\documentclass[preprint,aps,prd,amsmath,amssymb]{revtex4}
\usepackage{graphicx}
\usepackage{color}
\usepackage[latin1]{inputenc}
\usepackage{ulem}
\newcommand{\be}{\begin{eqnarray}}
\newcommand{\beq}{\begin{equation}}
\newcommand{\eeq}{\end{equation}}
\newcommand{\ee}{\end{eqnarray}}

\newcommand{\bmp}{\noindent\begin{minipage}{16cm}}
\newcommand{\emp}{\end{minipage}\vskip 7mm} 

\usepackage{graphicx}
\usepackage{dcolumn}
\usepackage{bm}
\usepackage{amsmath}
\usepackage{amsfonts}
\usepackage{bbm}
\usepackage{subfigure}

\def\drawbox#1#2{\hrule height#2pt
        \hbox{\vrule width#2pt height#1pt \kern#1pt
              \vrule width#2pt}
              \hrule height#2pt}

\def\Asym#1#2{\vcenter{\vbox{\drawbox{#1}{#2}
              \kern-#2pt 
              \drawbox{#1}{#2}}}}

\begin{document}
\title{WIMP Annihilation and Cooling of Neutron Stars}
\author{Chris {\sc Kouvaris}}
\email{kouvaris@nbi.dk}
 \affiliation{CERN Theory Division, CH-1211 Geneva 23,
 Switzerland, \\
University of Southern Denmark, Campusvej 55, DK-5230 Odense,
Denmark and \\
The Niels Bohr Institute, Blegdamsvej 17, DK-2100 Copenhagen,
Denmark   }

\begin{flushright} {\rm \small CERN--PH--TH/2007--145}
\end{flushright}
\par \vskip .05in


\begin{abstract}

We study the effect of WIMP annihilation on the temperature of a
neutron star. We shall argue that the released energy due to WIMP
annihilation inside the neutron stars, might affect the
temperature of stars older than 10 million years, flattening out
the temperature at $\sim 10^4$ K for a typical neutron star.

\end{abstract}


\maketitle

\section{Introduction}

Since Zwicky proposed the problem of the ``missing mass'' in 1933,
a lot of theoretical and experimental effort has been made in
order to unveil the nature of dark matter. Today, WMAP has
provided very accurate data regarding the matter density in the
universe~\cite{Spergel:2006hy}. The energy density of the universe
is composed of $4\%$ atoms and roughly $22\%$ dark matter. Data
from recent observations indicate that dark matter cannot be
attributed more than  $20\%$ to dim objects like black holes,
brown dwarfs and giant planets~\cite{Alcock:2000ph}. From a
theoretical point of view, several candidates rise from different
theories, such as
neutralinos~\cite{Jungman:1995df,Bertone:2004pz}, Majorana
neutrinos, and lately technibaryons provided by theories that are
not ruled out by the electroweak precision
measurements~\cite{Sannino:2004qp,Hong:2004td,Dietrich:2005jn,Gudnason:2006ug,Gudnason:2006yj,Kouvaris:2007iq}.
From the experimental point of view, the focus is on the direct
and indirect detection of dark matter particles. The direct
detection might occur in underground experiments like CDMS that in
principle can detect recoil energies from collisions between
Weakly Interacting Massive Particles (WIMPs) and nuclei, or
atmospheric experiments like XQC, where strongly interacting
particles might collide with the detector. The indirect detection
 might occur via gamma-ray and neutrino telescopes, where
the presence of WIMPs can be detected indirectly, by observing
products of WIMP annihilations. In particular, provided that WIMPs
can annihilate and because they can be trapped inside the earth or
the sun, such annihilations would produce jets of particles and
more specifically neutrinos coming straight from the center of the
earth or the sun, that could be possibly detected by neutrino
telescopes~\cite{Press:1985ug,Gould:1987ir,Gould:1987ww}. On the
other hand, gamma-ray telescopes can in principle detect
gamma-rays produced by WIMP annihilation at the center of the
galaxy~\cite{Zeldovich:1980st}. Both direct and indirect detection
experiments can impose strong constraints on the cross section of
the WIMP with the nuclei. For instance, heavy Dirac neutrinos have
been excluded as WIMPs for masses up to several TeV, because their
elastic cross section with nuclei is sufficiently large and
therefore they should have been detected by now in
CDMS~\cite{Akerib:2004fq}.

In this paper we investigate the possibility of a different kind
of indirect signature of WIMP annihilation. Instead of looking at
the indirect signals from the annihilation of trapped WIMPs inside
the earth or the sun, we examine the consequences of WIMP
annihilation on the temperature of neutron stars. The neutron
stars are massive compact objects with very low temperatures.
 Naively one
might expect that since the mass of the trapped WIMPs inside a
neutron star represents a tiny fraction of the overall mass of the
star, such an effect should be negligible. However, the
annihilation of massive particles inside the star releases a huge
amount of energy that is heating up the star. As we shall argue,
once the accretion rate of dark matter particles equilibrates the
rate of annihilation, the amount of released energy is independent
of the star's temperature and therefore at late times the WIMP
annihilation can keep the star at a constant temperature that
depends on the mass and the radius of the star, the cross section
of annihilation and the local dark matter density of the star.

The paper is organized as follows: First we calculate the rate of
dark matter accretion onto the neutron star including general
relativity corrections. Then we calculate the annihilation rate
for the WIMPs and we calculate the effect of the WIMP annihilation
on the cooling curves of a typical neutron star made of regular
nuclear matter. We present our conclusions in the last section.

\section{WIMP's Accretion Rate Onto the Neutron Star}

The accretion of dark matter particles inside the earth and the
sun is not a new subject. Press and Spergel studied first
in~\cite{Press:1985ug} the capture rate of WIMPs inside the earth
and the sun. More elaborate calculations were also done by Gould
~\cite{Gould:1987ir,Gould:1987ww}, taking into account several
effects specifically for the case of the earth and the sun. An
estimate of the accretion rate onto a neutron star was also
provided by Goldman and Nussinov~\cite{Goldman:1989nd}, who were
the first to study effects of WIMPs on neutron stars. In this
section we calculate the accretion rate of WIMPs onto a typical
neutron star including also general relativity corrections that
turn out to affect up to 70$\%$ the rate. Our derivation is along
the lines of~\cite{Press:1985ug}. We assume that the WIMP
population has a Maxwell-Boltzmann distribution of velocities \beq
p(v)dv=n_0\left(\frac{3}{2\pi\bar{v}^2}\right)^{3/2} 4 \pi
v^2\exp\left(\frac{-3v^2}{2\bar{v}^2}\right)dv, \eeq where
$\bar{v}=270\text{km}/\text{s}$, and $n_0$ is the number density
of the WIMPs in the neighborhood of the neutron star. The flux of
WIMPs (number per area per time) that crosses a spherical surface
of
 radius $R$ with velocity between $v$ and $v+dv$ and angle
 with respect to the normal between $\theta$ and $\theta +d\theta$, is
 \beq dF=n_0\left(\frac{3}{2\pi\bar{v}^2}\right)^{3/2}  \pi
v^3\exp\left(\frac{-3v^2}{2\bar{v}^2}\right)d(\cos^2\theta)dv.
\eeq We can express the flux in a more convenient way with respect
to the two invariants of the motion, i.e. the energy of the WIMP
per mass $E=(1/2)v^2$ and the angular momentum per mass
$J=vR\sin\theta$. The total accretion rate (number of particles
per time) is~\cite{Press:1985ug} \beq d\mathcal{F}=4\pi R^2dF=
n_0\left(\frac{3}{2\pi\bar{v}^2}\right)^{3/2}
\exp\left(\frac{-3E}{\bar{v}^2}\right)4\pi^2dEdJ^2.\label{flux}
\eeq The actual capture rate of WIMPs by the star can be
calculated in two steps. The first one is to determine what part
of the phase space for $E$ and $J$ can give orbits for the WIMPs
that intersect with the neutron star. In the second step we have
to determine what fraction of the particles that intersect with
the star, lose enough energy so they can be trapped inside the
star. For the first part of the calculation, we have to find the
trajectories that have a perihelion (closest distance to the
center of the star) at most equal to the radius of the star. Press
and Spergel calculated this using classical Newtonian mechanics.
The perihelion is \beq
r_{\text{peri}}=\left(\frac{J^2}{GM}\right)\big{/}
\left(1+\sqrt{1+2\frac{J^2}{GM}\frac{E}{GM}}\right), \eeq where
$G$ is the gravitational constant and $M$ is the mass of the
neutron star. This expression has two limiting cases. For
$J^2E<<(GM)^2$ \beq r_{\text{peri}}=\frac{J^2}{2GM},\label{hard}
\eeq and for $J^2E>>(GM)^2$ \beq
r_{\text{peri}}=\sqrt{\frac{J^2}{2E}}. \eeq The two regimes are
separated by the hyperbola \beq \frac{J^2}{GM}\frac{E}{GM}=1. \eeq
Since Eq.~(\ref{flux}) falls exponentially with respect to the
energy, we approximate (as it is done in~\cite{Press:1985ug}) the
exponential as unity with $E$ varying from zero to
$(1/3)\bar{v}^2$, which is the characteristic scale of the
exponential. In addition, $E$ is also restricted to values smaller
than $E_0$, where $E_0$ represents a constant that parametrizes
the maximum kinetic energy per mass of the WIMP at asymptotically
large distance from the star in order for the WIMP to be captured
by the star. We shall determine $E_0$ later on. Therefore, as it
was argued in~\cite{Press:1985ug}, the accretion rate of
capturable WIMPs is given by (\ref{flux}), if we integrate over
$E$ from zero to the minimum between $(1/3)\bar{v}^2$ and $E_0$
and over $J^2$ from zero up to $2GMR$ (that comes
from~(\ref{hard}) if
 $r_{\text{peri}}=R$), where $R$ is now the radius of the star.
The rate can be written as \beq \mathcal{F}=n_0 \left(
\frac{3}{2\pi\bar{v}^2}\right)^{3/2}
4\pi^2(2GMR)~\text{min}\left(\frac{1}{3}\bar{v}^2, E_0\right).
\label{capture}\eeq This formula differs by a factor of 2 with
respect to the corresponding one in~\cite{Press:1985ug}, as it was
first pointed out by Gould~\cite{Gould:1987ir}. Although the above
formula is a good approximate relation for the capture rate of
WIMPs for the sun and the earth, for the case of a neutron star,
general relativity corrections increase the rate significantly. To
show this, we are going to use the timelike geodesic equations
that describe the motion of a particle in a Schwarzschild metric.
The trajectory for nonrelativistic particles (as the WIMPs) is
given by~\cite{Collins:1973xf} \beq
\left(\frac{du}{d\phi}\right)^2 =2mu^3-u^2
+\frac{2}{J^2}mu+\frac{2E}{J^2}, \label{trajectory} \eeq where
$m=GM$ (in natural units) and $u=1/r$. We want to find for what
values of the phase space of $E$ and $J^2$, the perihelion becomes
smaller or equal to the radius of the neutron star. At the
perihelion $du/d\phi=0$ and $u=1/R$ (for $r_{\text{peri}}=R$). If
we express $E$ in units of $GM/R$ and $J^2$ in units of $GMR$,
Eq.~(\ref{trajectory}) gives \beq
E=\frac{1}{2}\left(1-\frac{2GM}{R}\right)J^2-1. \label{grperi}\eeq
The above equation gives the relation between $E$ and $J$ in order
the perihelion to be $R$. For $E=0$, $J^2=2/(1-2GM/R)$. This means
that the allowed phase space for $J^2$ has increased compared to
the Newtonian case from 2 (in units again of $GMR$) to
$2/(1-2GM/R)$. For a typical neutron star of mass 1.4 the solar
mass $M_{\odot}$ and a radius of 10 km, $J^2=2/(1-2GM/R)\simeq
3.4$. This a $70\%$ increase in the phase space of $J^2$ and the
capture rate compared to the classical case. Therefore
Eq.~(\ref{capture}) should be modified as \beq \mathcal{F}=n_0
\left( \frac{3}{2\pi\bar{v}^2}\right)^{3/2}
4\pi^2(2GMR)\frac{1}{1-2GM/R}~\text{min}\left(\frac{1}{3}\bar{v}^2,
E_0\right). \label{capture2} \eeq

Now we estimate $E_0$. We shall show that in the case of a neutron
star, $E_0>>(1/3)\bar{v}^2$ and therefore Eq.~(\ref{capture2})
should be always taken with $(1/3)\bar{v}^2$ as the minimum. A
WIMP that intersects with the neutron star might or might not
interact with the nuclear matter inside the star. If it does
scatter at some point, the recoil energy and the energy loss of
the particle is $0<T<4m_nm_{\chi}/(m_n+m_{\chi})^2$, where $T$ is
the recoil energy and $m_n$ and $m_{\chi}$ are the masses of the
nucleus and the mass of the WIMP respectively. If we assume that
the scattering is isotropic, then the recoil energy should be
uniformly distributed. The condition that holds in order to
capture a WIMP is that the energy loss in the scattering should be
at least equal to the initial kinetic energy of the WIMP at
asymptotic large distance from the star. If this condition is
fulfilled, the WIMP stays in a bound state with the star.
Therefore, for an average collision that takes place in the star,
this condition can be written as \beq \Delta E=
\frac{2m_nm_{\chi}}{(m_n+m_{\chi})^2}\left(1-\sqrt{1-\frac{2GM}{R}}\right)\geq
E_0, \eeq where we took into account the gravitational redshift
effect. Again we have chosen to set $c=1$. If we plug the typical
values we used before for the mass and the radius of a neutron
star and a WIMP mass of the order of TeV, we find that $E_0$ is
three orders of magnitude larger than $(1/3)\bar{v}^2$ and
therefore for all the cases of interest, we can
use~(\ref{capture2}) with $(1/3)\bar{v}^2$ as the minimum.

Eq.~(\ref{capture2}) gives the rate of capturable WIMPs, that is
the number of WIMPs per second that intersect with the neutron
star. However, as mentioned before, in order for the WIMP to be
trapped in the star, one or more collisions have to take place. We
know from classical mechanics that if the WIMP does not scatter
while travelling through the star, it cannot be captured by the
star. We calculate now the fraction of the capturable WIMPs (given
by~(\ref{capture2})), that can yield scatters into bound orbits.
Such a fraction would depend strongly on the elastic scattering
cross section of the WIMP-nucleus system. For the typical neutron
star of mass $1.4M_{\odot}$ and $R=10$ km, the average density of
neutrons is $\rho = 3M/(4\pi R^3m_n)\simeq 4 \times 10^{38}$
neutrons$/\text{cm}^3$. If we take a typical value for the elastic
cross section between WIMP-neutron of the order of
$10^{-44}$~$\text{cm}^2$, the mean free path is about 1 km. Since
for an average WIMP, even one scattering is enough to result in a
bound orbit around the star, a mean free path of 1 km means that
the fraction of the capturable WIMPs that will be trapped is very
close to 1. To entertain this, if for simplicity we assume that
the WIMP has a straight trajectory while inside the neutron star,
a segment of 1 km corresponds to an impact parameter of 9.9 km,
which means that only if the WIMP intersects between 9.9-10 km
from the center of the star will travel a distance less than 1 km
inside the star. Obviously if the cross section is larger than
$10^{-44}$ $\text{cm}^2$, the fraction saturates even faster to 1.
We shall give now a more quantitative answer about the fraction
following the derivation of~\cite{Press:1985ug}. The fraction $f$
of the particles that undergo one or more scatterings while inside
the star is defined as \beq f=\Big{\langle} 1-
\exp\Big{[}-\int\frac{\sigma_{\chi}\rho}{m_n}dl\Big{]}\Big{\rangle}\simeq
\Big{\langle}\int\frac{\sigma_{\chi}\rho}{m_n}dl \Big{\rangle},
\label{f1}\eeq where the last approximation holds if the elastic
cross section between WIMP-nucleus $\sigma_{\chi}$ is smaller than
$\sigma_{\text{crit}}=m_nR^2/M \simeq 6 \times
10^{-46}~\text{cm}^2$. Eq.~(\ref{f1}) now reads \beq f \simeq
\frac{\sigma_{\chi}}{\sigma_{\text{crit}}}\Big{\langle}\int\frac{\rho}{M/R^3}\frac{dl}{R}
\Big{\rangle}. \label{f}\eeq In order to find $f$, we examine
trajectories with $E=0$ as in~\cite{Press:1985ug}, since
$E<<GM/R$. We average over $J^2$ (that ranges from 0 to 3.4 for
our typical neutron star). For an accurate calculation of $f$, we
need to know the exact density profile of the star, in order to
know explicitly the mass of the star $M(r)$ as a function of the
radius.
Here we give an estimate of $f$, by assuming for simplicity that
the density of the star is constant through the whole volume. This
means that $M(r)/M=(r/R)^3$ (where $M$ is the total mass of the
star). If we take the derivative of Eq.~(\ref{trajectory}) with
respect to $\phi$, we get the following equation of motion for the
WIMP inside the neutron star \beq
\frac{d^2\hat{u}}{d\phi^2}+\hat{u}=\frac{1}{J^2}\frac{M(r)}{M}+3\frac{GM}{R}\frac{M(r)}{M}\hat{u}^2=
\frac{1}{J^2\hat{u}^3}+\frac{3GM}{R\hat{u}},\label{diffe}\eeq
where $\hat{u}=Ru$ and again $J^2$ is measured in units of $GMR$.
The initial condition of this differential equation are
$\hat{u}(\phi=0)=1$, which means that we have chosen $\phi=0$ at
the point where the WIMP crosses the surface of the star. For the
velocity \beq
\frac{d\hat{u}(\phi=0)}{d\phi}=\sqrt{\frac{2GM}{R}-1+\frac{2}{J^2}}.
\eeq The length of the path of the particle travelling inside the
star is \beq
\frac{dl}{R}=\frac{d\phi}{\hat{u}^2}\sqrt{\left(\frac{d\hat{u}}{d\phi}\right)^2+\hat{u}^2}\label{dl}.
\eeq We can find the length of the path of a particle inside the
star if we integrate $\phi$ from 0 up to the angle that
$\hat{u}(\phi)=1$ again, which is the point where the particle
exits from the star.
 Using Eqs.~(\ref{f}),~(\ref{dl}), and after having solved Eq.~(\ref{diffe})~numerically,
 and averaging over $J^2$ from 0 to 3.4,
  we found that  the average path inside the star is $1.87R$ and
  $f=0.45\sigma_{\chi}/\sigma_{\text{crit}}$. We should emphasize
  that the above estimate of $f$ holds for
  $\sigma_{\chi}<\sigma_{\text{crit}}$. It is understood that if
  $\sigma_{\chi}>\sigma_{\text{crit}}$, $f$ increases, saturating
  to 1 as soon as $\sigma_{\text{crit}}$ becomes larger than
  roughly $10^{-45}$ $\text{cm}^2$. In addition, we should mention
  that our estimate is a lower bound for $f$. This is because in
  our derivation we assumed a constant density. Neutron stars are expected to be
  denser as we approach the
  center. In this case, $f$ will be larger than our estimate. To illustrate
  this, we examined the extreme case where all the mass of the
  star is concentrated at the core. In this scenario, although
  somewhat unrealistic, $f$ increases drastically because for a large range
  of $J^2$, the orbits follow
  spirals around the center. A large fraction of
  the particles will be trapped in the star not because of
  energy loss due to a collision, but due to the fact that they are trapped
  gravitationally. Our expectation is that the real case should
  be somewhere in the middle and therefore we consider our
  previous estimate for $f$ as a lower bound.

Using Eq.~(\ref{capture2}) and the values for our typical neutron
star $M=1.4M_{\odot}$, $R=10$ km, we get the rate of accretion of
dark matter inside the star in particles per second \beq
\mathcal{F}=\frac{3.042 \times 10^{25}}{m_{\chi}
(\text{GeV})}\times A \times f, \eeq where $A$ is a constant that
parametrizes the local dark matter density in the vicinity of the
neutron star in units of 0.3 GeV$/\text{cm}^3$ (which is the
standard dark matter density around the earth). For cross sections
$\sigma_{\chi}>10^{-45}$ $\text{cm}^2$, $f=1$, otherwise $f$ is
given by $f=0.45\sigma_{\chi}/\sigma_{\text{crit}}$.

\section{Annihilation rate}

Once the WIMP undergoes one scattering inside the star, it loses
on the average enough energy to be captured by the star. Even if
the kinetic energy is sufficient enough to make it exit from the
star, it will be forced to return and probably scatter again
loosing even more energy. The WIMP can repeat this process several
times until its kinetic energy reduces down to the thermal
velocity inside the star. It is easy to show that for most cases
of interest, the WIMP thermalizes very fast compared to the other
time scales of the problem. We can make a very rough estimate of
how long it takes for a WIMP moving with the average velocity of
270 km$/\text{sec}$ to obtain a velocity comparable to the thermal
velocity. Let's assume that the WIMP has undergone one scattering
and therefore has lost on average energy of
$(2m_n/m_{\chi})v_{\text{esc}}^2/2$, where $v_{\text{esc}}$ is the
escape velocity from the star. A simple approximate calculation
shows that it will take a few seconds before the WIMP intersects
again with the star and loose again a fraction $2m_n/m_{\chi}$ of
its energy. This depends on the cross section of elastic collision
between WIMP-nucleus, but as we showed in the previous section,
for $\sigma_{\chi}>10^{-45}$ $\text{cm}^2$, this will happen on
average. In that case, it will take just a few hours before the
kinetic energy of the WIMP reduces down to the thermal velocity.
So our conclusion is that for not extremely small cross sections,
captured WIMPs will thermalize pretty fast and they will have a
Maxwell-Boltzmann distribution in velocity and in distance from
the center of the star.

The population of dark matter WIMPs inside the star is governed in
principle by three processes. The first one is the accretion of
WIMPs onto the star. The second is the evaporation and the third
one is the annihilation. Once the WIMPs thermalize inside the
star, they follow a Maxwell-Boltzmann distribution in velocity.
Particles that are in the tail of the distribution (with large
velocities) can escape from the star, if the velocity is larger
than the escape velocity of the star. However, in the cases we are
interested in, this process is exponentially suppressed. The rate
of evaporating particles is proportional to
$\text{exp}(-GMm_{\chi}/RT)$~\cite{Krauss:1985aa}. Since the
radius of a neutron star is very small, and we are interested in
WIMPs with mass of the order of TeV, the rate becomes negligible.
For a temperature of 100 keV (which corresponds to a typical
temperature of a neutron star that is a few thousand years old),
and for $m_{\chi}=1$ TeV, the suppression is $\text{exp}(-10^7)$.
Therefore we can safely ignore the evaporation process.

If the WIMP is a Majorana particle, for example a Majorana
neutrino, it is possible to co-annihilate with another one. The
annihilation depends on the cross section as well as the density
of the WIMPs inside the star. The annihilation cross section
should not be confused with the elastic cross section between
WIMP-nucleus that was mentioned before. One big difference between
the two is that the annihilation cross section for Majorana
particles is usually velocity dependent. If the WIMP is a Majorana
neutrino, it has an elastic cross section with nuclei \beq
\sigma_{\chi}=\frac{2G_F^2}{\pi}\mu^2I_s, \eeq where $\mu$ is the
reduced mass of the system WIMP-nuclei and $I_s$ is a form factor
that depends on the nuclei~\cite{Lewin:1995rx}. The annihilation
cross section of two Majorana neutrinos depends on what channels
are open for annihilation. If the mass of the Majorana neutrinos
is larger than 100 GeV, the dominant channel is the annihilation
to a pair of $W^+-W^-$ mediated by a $Z$
boson~\cite{Enqvist:1990yz}. In the case where $m_{\chi}>>100$
GeV, the annihilation cross section is given by \beq
\sigma_A=\frac{G_F^2m_{\chi}^2}{3\pi}\beta^2,
\label{crossanni}\eeq where $\beta$ is the velocity of the WIMP
(at the center of mass frame). Once the WIMP will get thermalized
in the star, $\langle\beta^2\rangle=3T/(2m_{\chi})$ ($T$ being the
temperature in the star).

The annihilation rate of WIMPs in the star is given by \beq
\Gamma_A=\langle\sigma_{\chi}v\rangle \int n^2dV, \eeq where
$\langle\sigma_{\chi}v\rangle$ is the thermally averaged
annihilation cross section times the velocity and $n$ is the
density of the WIMPs inside the neutron star. For convenience we
shall assume that the density is constant inside the star. In this
case, the population of WIMPs inside the star is governed by \beq
\frac{dN}{dt}=\mathcal{F}-C_AN^2. \label{eqpopu}\eeq The constant
$C_A=\langle\sigma_{\chi}v\rangle/V$, where $V$ is the volume of
the star. The accretion rate $\mathcal{F}$ was derived in the
previous section. The solution of Eq.~(\ref{eqpopu}) is \beq
N(t)=\sqrt{\frac{\mathcal{F}}{C_A}}\tanh(\frac{t}{\tau}).
\label{popu}\eeq The time scale $\tau=1/\sqrt{\mathcal{F}C_A}$.
The released energy due to the annihilation is \beq
E=C_AN^2m_{\chi}=\mathcal{F}\tanh^2(t/\tau)m_{\chi}.
\label{energyt} \eeq The amount of the released energy depends on
the time scale $\tau$. If $\tau$ is large compared to the age of
the known neutron stars, the hyperbolic tangent is suppressed and
the effect of the dark matter on the temperature of the star is
negligible. For particles that the annihilation cross section is
velocity independent, $\tau$ is given by \beq \tau=\frac{2.1
\times 10^3~\text{years}}{\sqrt{\frac{Af\sigma_{39}}{m_{\chi}}}},
\eeq where $\sigma_{39}$ is defined through $\langle \sigma_{A}v
\rangle = \sigma_{39} 10^{-39}~\text{cm}^2$. In the case of a
Majorana particle, Eq.~(\ref{popu}) holds only approximately
because the cross section is velocity dependent and therefore in
thermal equilibrium temperature dependent. Since the temperature
of the star changes as a function of time, this means that there
is additional time dependence on the annihilation rate. Generally,
for a Majorana neutrino with mass larger than 100 GeV, the cross
section is given by Eq.~(\ref{crossanni}) times a factor
$\sin^4\theta$ which denotes the suppression of the cross section
due to a mixing between the left handed neutrino (that interacts
weakly) and a sterile right handed
neutrino~\cite{Enqvist:1990yz,Kouvaris:2007iq}. The time scale
$\tau$ is \beq
\tau=\frac{1}{\sqrt{\mathcal{F}C_A}}=\frac{5.98\times 10^5
\text{years}}{\sqrt{\frac{Af\sigma_{39}}{m_{\chi}^2}\left(\frac{T}{10^8}\right)}},
\label{tau1} \eeq where the temperature $T$ is measured in Kelvin
degrees and $\sigma_{39}$ is defined by the relation
$\sigma_A=\sigma_{39}10^{-39}\beta^2$ $\text{cm}^2$. The mass
$m_{\chi}$ is measured in GeV. Eq.~(\ref{tau1}) can be written
more conveniently in terms of the mixing angle $\sin\theta$ as
\beq \tau=\frac{2.52\times 10^5 \text{years}}{\sqrt{Af\sin^4\theta
\left(\frac{T}{10^8}\right)}}.
 \label{tau2} \eeq If the Majorana neutrino is exclusively left handed,
$\sin\theta=1$ and for a temperature of $10^8$ Kelvin, the time
scale is about $10^5$ years (depending on how much larger is the
local dark matter density in the vicinity of the star compared to
the one of the earth). As it can be seen from~(\ref{energyt}), the
annihilation saturates to $\mathcal{F}$ for times larger than
roughly $3\tau$. As we already mentioned, since the temperature of
the star changes in time, Eqs.~(\ref{tau1})~and~(\ref{tau2}) are
approximate. We shall return to the question of how fast the
annihilation rate reaches the saturated value in the next section.

\section{Cooling and Heating the Neutron Star}

In this section we investigate the influence of the WIMP
annihilation on the temperature of the neutron star. Naively, one
would expect that such an effect should be negligible due to the
fact that the accretion of dark matter represents a tiny fraction
of the whole mass of the star. However, there are two elements
that make this investigation interesting. The first one is that
although the trapped dark matter represents a small fraction of
the mass of the star, the annihilation of two WIMPs releases a
huge amount of energy. After the annihilation, this energy is
carried mostly by leptons, quarks and photons. Since they cannot
escape from the star, they will heat it up. A small portion of the
energy will be carried by neutrinos that will escape and they will
not contribute to the heating of the star. However, to first
approximation, the energy carried by the neutrinos is negligible
compared to the one carried by quarks, leptons and photons. So, we
are going to assume that the whole annihilation energy will be
carried not by neutrinos and therefore this energy will heat up
the star. The second reason we investigate this effect is that the
energy released by the annihilation and consequently the
emissivity of this process does not scale with the temperature. As
long as equilibrium between accretion and annihilation has been
reached, the released energy per time remains unchanged. All the
dominant cooling processes of a neutron star scale with positive
powers of $T$. This means that inevitably, even if the emissivity
due to WIMP annihilation is small, as the temperature of the star
decreases, the WIMP annihilation emissivity will dominate at some
point.

Let's assume for the moment and we shall examine later under what
conditions this is possible, that the time $3\tau$ has been
reached and the released energy from the annihilation of the WIMPs
is $E=\mathcal{F}m_{\chi}$. The emissivity, i.e. released energy
per volume per time is \beq \epsilon_{\text{dm}}=\frac{E}{4 \pi
R^3/3}=\frac{3\mathcal{F}m_{\chi}}{4 \pi R^3}=A~1.16 \times 10^4
~\text{erg} ~\text{cm}^{-3} \text{s}^{-1}. \label{edm}\eeq

There are several processes that contribute to the cooling of a
neutron star depending on the form of matter, the temperature, and
the density of the star. If the star is sufficiently dense,
unpaired quark matter or other exotic phases might occur deeply
inside the star. If the neutron star has unpaired quark matter,
for the first million years cools very fast due to neutrino
emission via the direct Urca process. In this case the emissivity
scales as $\epsilon_{\nu} \sim T^6$. Direct Urca processes are
allowed in sufficiently dense nuclear matter, nuclear matter with
pion condensation, kaon condensation, or nonzero hyperon density,
and in all phases of quark matter except
CFL~\cite{Alford:2004zr}~and references therein. For neutron stars
that are not in sufficiently high density, direct Urca processes
$n \rightarrow p + e +\bar{\nu}$, $p+e \rightarrow n + \nu$ are
kinematically forbidden. In this case, a bystander neutron is
needed in order to assist kinematically the reaction. This is the
so-called modified Urca process. During the epoch dominated by the
modified Urca, the star loses energy through neutrino emission, by
converting protons and electrons to neutrons and vice versa. The
emissivity of this process scales as $\sim T^8$. It
is~\cite{Shapiro:1983du} \beq \epsilon_{\nu}=(1.2 \times 10^4
~\text{erg} ~\text{cm}^{-3} \text{s}^{-1}) \left( \frac{n}{n_0}
\right)^{2/3} \left( \frac{T}{10^7 \text{K}}\right)^8,
\label{murca}\eeq where $n$ is the baryon density of the star and
$n_0=0.17~\text{fm}^{-3}$ is the baryon density in nuclear matter.
In our calculation for the neutron star of $M=1.4M_{\odot}$ and
$R=10$ km, we are going to use the average density $n=3.9 \times
10^{38}$ particles per $\text{cm}^3$ and therefore $n/n_0=2.3$.

After the first million years and roughly as soon as the
temperature of the star drops below $10^8$ K, the dominant
mechanism of cooling is not anymore through neutrino emission, but
through photon emission from the surface of the star. The rate of
heat loss from the surface of the star is \beq L_{\gamma}=4 \pi
R^2 \sigma T^4_{\text{surface}}, \eeq where $\sigma$ is the
Stefan-Boltzmann constant and $T_{\text{surface}}$ is the
temperature of the surface of the star. The surface of a neutron
star is usually colder than the interior of the star. This change
in the temperature occurs inside the crust of the neutron star,
taking place within 100 meters below the surface. The surface
temperature of the star is well approximated
by~\cite{Gundmundsson1,Gundmundsson2,Page:2004fy} \beq
T_{\text{surface}}=(0.87 \times 10^6 \text{K}) \left(
\frac{g_s}{10^{14} \text{cm/s}^2} \right)^{1/4} \left(
\frac{T}{10^8 \text{K}} \right)^{0.55}, \eeq where $T$ is the
interior temperature of the star and $g_s=GM/R^2$ is the surface
gravity. The rate of heat loss $L_{\gamma}$ can now be expressed
in terms of the interior temperature as \beq L_{\gamma}=4 \pi R^2
\sigma (0.87 \times 10^6 \text{K})^4 \left( \frac{g_s}{10^{14}
\text{cm/s}^2} \right) \left( \frac{T}{10^8 \text{K}}
\right)^{2.2}. \eeq If we divide $L_{\gamma}$ over the volume of
the star, we can get an ``effective'' emissivity of photons
measured in energy over volume and time \beq
\epsilon_{\gamma}=\frac{L_{\gamma}}{(4/3) \pi R^3}= 1.8 \times
10^{14} \left(\frac{T}{10^8 \text{K}}\right)^{2.2}
\text{erg}~\text{cm}^{-3}~\text{s}^{-1}, \label{photon}\eeq where
we used $g_s=1.85 \times 10^{14}~\text{cm}/{s}^2$.

In order to be able to derive the temperature as a function of
time, we need to know the heat capacity of the star. For a gas of
noninteracting fermions, the specific heat is given
by~\cite{Shapiro:1983du}
 \beq c_V=\frac{k_B^2T}{3 \hbar^3 c} \sum_i
 p_F^i\sqrt{m_i^2c^2+(p_F^i)^2}, \eeq where the sum runs over the
 different species. In the case we investigate, namely the one of
 noninteracting nuclear matter, $i$ runs over $n$, $p$, $e$ and the Fermi momenta
 for neutral matter in weak equilibrium are
 \begin{eqnarray} & p_F^n &=(340~\text{MeV})\left(\frac{n}{n_0}\right)^{1/3} \\
 & p_F^p & =p_F^e=(60~\text{MeV})\left(\frac{n}{n_0}\right)^{2/3}.
\end{eqnarray}
\begin{figure}[!tbp]
  \begin{center}
    \mbox{
      \subfigure{\resizebox{!}{4.8cm}{\includegraphics{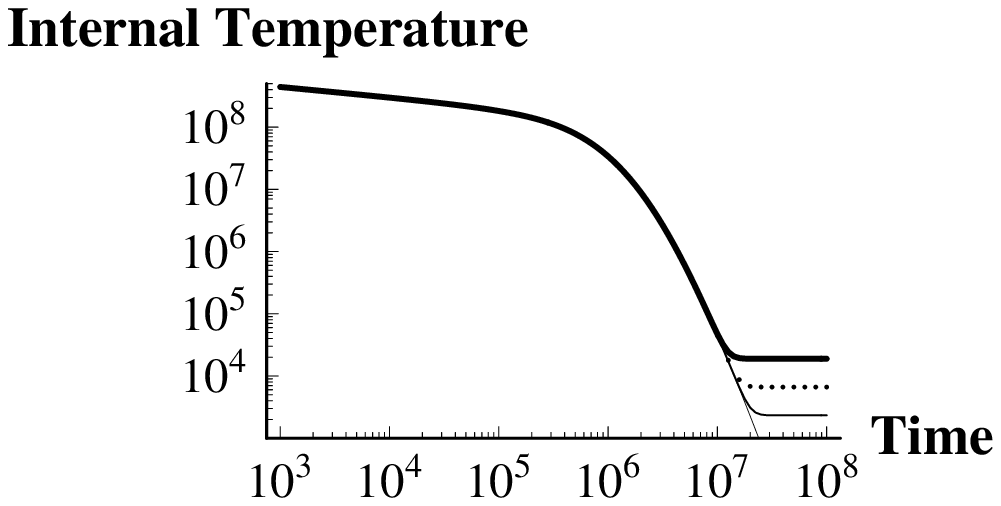}}} \quad
      \subfigure{\resizebox{!}{4.8cm}{\includegraphics{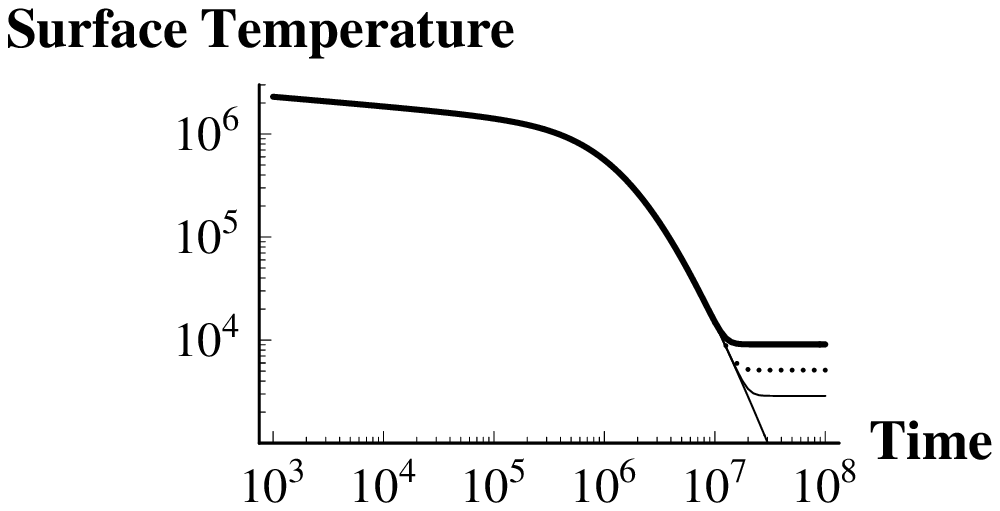}}}
      }
    \caption{\emph{Left Panel}: The internal temperature of a
    neutron star (in Kelvin) with $M=1.4M_{\odot}$ and $R=10$ km as a
    function of time (in years). The solid line that crosses the
    time axis corresponds to the case where the effect of dark
    matter annihilation is neglected. The thin solid line
    corresponds
    to a local dark matter density for the star of 0.3
    GeV$/\text{cm}^3$. The dashed and the thick solid lines
    correspond to local densities of 3 and 30 GeV$/\text{cm}^3$
    respectively.
     \emph{Right Panel}:As in the left panel for the surface temperature of the neutron star.}
    \label{fig:Omega}
    \end{center}
\end{figure}
The cooling of the star is dictated by the differential equation
\beq \frac{dT}{dt}=\frac{-L_{\nu}-L_{\gamma}+L_{\text{dm}}}{V
c_V}=\frac{V(-\epsilon_{\nu}-\epsilon_{\gamma}+\epsilon_{\text{dm}})}{V
c_V}=\frac{-\epsilon_{\nu}-\epsilon_{\gamma}+\epsilon_{\text{dm}}}{
c_V}, \label{cooling-eq}\eeq where the volume of the star $V$
drops out at the end. We have neglected the contribution of the
WIMPs to the specific heat, since they consist a tiny fraction of
the mass of the star. We solved Eq.~(\ref{cooling-eq})
numerically, by imposing an initial temperature for the star of
$10^{10}$ K at very early time. However, we should emphasize that
the temperature is very insensitive to the initial condition. It
affects only the temperature during the first years of the star's
life, but it has no effect later on. In Fig.~1, we have plotted in
a logarithmic scale the internal and the surface temperature of
the star as a function of time, starting from time $t=1000$ years,
up to 100 million years. We have plotted the temperature for 3
different cases that correspond to different local dark matter
densities for the vicinity of the star. We chose the local dark
matter densities to be the one of the earth (0.3
GeV$/\text{cm}^3$), 10 times larger and 100 times larger the one
of the earth. For comparison we also plotted the cooling curve of
the same star without including the effect of dark matter
annihilation. As we can see from the figures, the dark matter
annihilation does not affect the temperature of the star up to
$t=10$ million years. Between $10^3$ and $10^6$ years, the star
cools due to neutrino emission via the modified Urca process,
while after roughly 1 milion years through photon emission from
the surface of the star. However, by inspection of
Eqs.~(\ref{edm}),~(\ref{murca}),~and~(\ref{photon}), we see that
the dark matter annihilation emissivity scales with the lowest
power of $T$. More precisely, $\epsilon_{\text{dm}}$ does not
depend on $T$, once equilibrium between the rate of accretion of
dark matter and the rate of annihilation has been established.
This means that inevitably, when the temperature of the star drops
sufficiently, the power of the dark matter annihilation that heats
up the star will equate the power of photon emission and as a
result the temperature will remain flat as a function of time.
This happens roughly at $t=10$ million years and at surface
temperatures between 3000 to 10000 K (depending on the local dark
matter density, the mass and the radius of the star). To entertain
the possibility of having a neutron star with a local dark matter
density 10 or 100 times larger than 0.3 GeV$/\text{cm}^3$, we can
use a indicative profile density for the dark matter halo. We
consider the Navarro-Frenk-White profile, where the dark matter
density is given by~\cite{Navarro:1996gj}\beq \rho
(r)=\frac{\rho_0}{\left(\frac{r}{R}\right)^{\gamma}\left(
1+\left(\frac{r}{R}\right)^{\alpha}\right)^{\frac{(\beta-\gamma)}{\alpha}}}.
\eeq This profile has a spike in the center of the galaxy for
positive $\gamma$. We shall use $\alpha=1$, $\beta=3$, $\gamma=1$,
$R=20$ kpc and $\rho_0=0.235$ GeV$/\text{cm}^3$. Given this
density profile, a neutron star with density 10 and 100 times
larger than the local dark matter density of the earth, should be
1.37 and 0.15 kpc from the center of the galaxy respectively. The
position of the earth is roughly 8 kpc from the center of the
galaxy. This means that a neutron star that exhibits the flatness
in temperature of $\sim10^4$ K for time $t>10$ million years due
to the dark matter annihilation, should lie at least $\sim 6.5$
kpc away from the earth. This limit can be improved, if the star
has ``more convenient'' mass and radius from what we have
considered for a typical neutron star. The emissivity
$\epsilon_{\text{dm}}$ is proportional to $\mathcal{F}$, which is
proportional to the factor $MR/(1-2GM/R)$. The emissivity
$\epsilon_{\text{dm}}$ is also proportional to the local dark
matter density. When we quote results with local dark matter
density 10 times larger the one of the earth, this does not imply
per se that the star has to be at a region of the galaxy with
$\rho_{\text{dm}}=3$
 GeV$/\text{cm}^3$, but the factor $MR/(1-2GM/R)$ times the local
 dark matter density should be ten times the same factor for our
 typical star with $M=1.4M_{\odot}$ and $R=10$ km times the
 density of 0.3 GeV$/\text{cm}^3$. For example, a neutron star of
 $M=2M_{\odot}$ and $R=6$ km, gives a factor of $\sim 26.5$ compared
 to our typical star. Therefore our results for density 100 times
 the density of the earth are applied also for a star with mass
 and radius given in the previous sentence and a local dark matter
 density of only $100/26.5=3.77$ times the density of dark matter around the earth.

Now we return to answer a question we posed earlier. In the
results we have presented, we have assumed that equilibrium
between the accretion and the annihilation rate has taken place
before $t=10$ million years, where the effect of the WIMP
annihilation becomes important. If $\tau$ is much larger than 10
million years, the annihilation rate has not reached the
asymptotic value $\sim \mathcal{F}$ and the effect on the
temperature of the star will be negligible. We saw in Fig.~1 that
even if the asymptotic annihilation rate is reached very early,
the WIMP annihilation does not affect the temperature for $t<10$
million years (or equivalently for temperatures higher than $10^4$
K). This means that up to $10^7$ years, the temperature of the
star is controlled by modified Urca and photon emission. We
already mentioned that if the WIMP annihilation cross section is
velocity dependent (like in the case of Majorana neutrinos),
Eqs.~(\ref{energyt}),~(\ref{tau1}),~and~(\ref{tau2}) hold only
approximately since $\tau$ is temperature dependent (and
implicitly time dependent). In particular,
Eqs.~(\ref{tau1}),~and~(\ref{tau2}) can give a lower bound
estimate of how fast the asymptotic annihilation rate is reached.
For most candidates of our interest, Majorana neutrinos, or
Majorana technibaryons~\cite{Kouvaris:2007iq}, we can safely take
$f=1$, since the elastic cross section with nuclei is not smaller
than $10^{-45}$ $\text{cm}^2$. By using Eq.~(\ref{tau2}), with
$f=1$, $A=10$, and an ``average'' temperature of $10^8$ K for the
star, we get that for cross section suppression $\sin\theta>0.1$,
$\tau \sim 10^6$ years. In reality, the situation is better,
because this estimate is done assuming constant $T=10^8$ K.
However, the temperature of the star is much higher at the
beginning ($\sim 10^{10}$ K) and due to modified Urca drops down
to $T=10^8$ K roughly at $t=10^6$ years. The higher temperature at
the beginning, brings the annihilation rate faster to the
asymptotic value. In addition, for $10^6<t<10^7$ years, where the
star cools mainly due to photon emission, the annihilation rate
can also improve towards the asymptotic value, although since the
temperature falls really fast, this happens with a slow rate. For
the two cases we mentioned, Majorana neutrinos and Majorana
technibaryons, the annihilation rate reaches the asymptotic value
before 10 million years. For example for a majorana neutrino,
Eq.~(\ref{tau2}), for $\sin\theta=1$, $A=100$, and $f=1$, gives
$2.52 \times 10^4$ years (for $T=10^8$ K). For a Majorana neutrino
(or Majorana technibaryon) with suppressed coupling to the $Z$
boson in order to account for the right dark matter density, for a
mass of 1 TeV, $\sin\theta= 0.26$~\cite{Kouvaris:2007iq}. In this
case, Eq.~(\ref{tau2}) gives a characteristic time scale $\tau=3.7
\times 10^5$ years, with $A$, $f$, and $T$ as before. Therefore,
for most cases of interest and unless the cross section is very
small, the annihilation rate reaches the accretion rate before 10
million years, which is the time where the dark matter
annihilation affects the temperature of the neutron star.

\section{Conclusions}

We investigated the effect of WIMP annihilation on the temperature
of a neutron star. We found that for a typical neutron star with a
local dark matter density at least 3 GeV$/\text{cm}^3$, and if the
WIMP has an elastic cross section not smaller than
$10^{-46}~\text{cm}^2$, the WIMP annihilation flattens out the
temperature of the star around $\sim 10^4$ K at $t=10$ million
years. Two neutron stars that have different local dark matter
densities would have different final temperatures scaling as $T
\sim \rho_{\text{dm}}^{1/2.2}$. Given the uncertainty in our
knowledge of the age of a neutron star and the fact that the peak
of a blackbody spectrum of $10^4$ K lies in the infrared, it is a
challenge to observe such an effect, which would possibly be a
signature of WIMP annihilation. Alternatively, instead of trying
to spot neutron stars with such a low temperature, it might be
more efficient to study pulsars, detected already by their
nonthermal emission and to constraint their thermal emission
putting an upper bound on their temperature.

We disregarded in our analysis the interesting possibility of
having an exotic phase of quark matter in the typical neutron
star. Exotic quark matter phases like color superconductivity can
have an effect on the cooling of a neutron
star~\cite{Alford:1997zt,Alford:1998mk,Rajagopal:2000wf,Alford:2003eg}.
 This could change the dominant cooling process for the first
million years, but it is highly unlikely that our conclusions
would change regarding the effect of the WIMP annihilation. We
also neglected reheating mechanisms that
 might possibly be present for old neutron stars. These mechanisms
 can be viscous dissipation of rotational energy within the
 star~\cite{alpar}, energy release due to weak deviations from
 beta equilibrium~\cite{Reisenegger:1994be}, accretion from
 interstellar gas, or
 other~\cite{Yakovlev:2004iq}~and references therein. All of these
 mechanisms are model dependent and it is not clear what is their
 effect on the temperature of an old neutron star compared to the WIMP
 annihilation mechanism. As a rule, in a nonsuperfluid old star without a
 magnetic field and an accreted envelope, the alternative
 mechanisms mentioned above cannot probably prevent the star from
 getting a very low temperature. In this case, WIMP annihilation
 might be the dominant reheating mechanism of the star.

Much theoretical work remains to be done. An interesting question
is to study if the effect of WIMP annihilation inside a neutron
star can change the suggested mechanisms for superbursts and
gravitational wave bursts proposed
in~\cite{Hong:2001gt,Berezhiani:2002ks,Lin:2005zd}. These
mechanisms assume a phase transition to quark matter at the center
of the neutron star via metastable phases, that can lead to a
burst. The released energy from the WIMP annihilation inside the
star might work as a catalyst, accelerating the phase transition,
making the metastable phases impossible to persist.

 \acknowledgments I am
grateful to Mark Alford, Maxim Khlopov, and Krishna Rajagopal for
their comments.
 The work of C.K. is supported by the Marie Curie Excellence
Grant under contract MEXT-CT-2004-013510.

\end{document}